\documentclass[a4,11pt]{cip-submit}  
\usepackage{afterpage}

\usepackage{mathptmx}
\usepackage{hyperref}
\hypersetup{
	colorlinks=false,
	pdfborder={0 0 0}
}
\def\Mr{\uppercase}

\usepackage{float}
\usepackage{wasysym}
\usepackage{graphicx,wrapfig,tikz}
\usepackage{caption}
\usepackage{subcaption}
\usepackage{amsmath,amsxtra,amssymb,latexsym,amscd,color,bm}
\usepackage[mathscr]{eucal}
\usepackage{epstopdf}
\usepackage{array}
\usepackage{cite}

\def\vsm{\vskip0.1cm}
\def\doi#1{\href{http://dx.doi.org/#1}{doi:#1}}

\def\titles#1{\title{\large\bf\noindent #1}}
\def\authors#1{\author{\begin{flushleft}{#1}\end{flushleft}}}
\def\authord#1#2{\indent\Mr{#1}\\
	\textit{\indent#2}\vsm}

\def\email#1{\bigskip\href{mailto:#1}{\textit{E-mail:}~{#1}}\\[3mm]}

\def\received#1{\vsm\textit{\indent Received #1}}
\def\accepted#1{\vsm\textit{Accepted for publication~#1}}

\def\and{$\text{\tiny AND }$}

\begin{document}
	\Year{2015}
	\Page{1}\Endpage{9}
	\titles{Testing the $f(R)$-theory of gravity}

\authors{\authord{Nguyen Anh $\mbox{Ky}^{1\ast}$,
		Pham Van  $\mbox{Ky}^{2\dagger}$,
		Nguyen Thi Hong $\mbox{Van}^{1,3\ddagger}$  \vspace*{2mm}}
	{$^1$Institute of physics, Vietnam academy of science and technology, 10 Dao Tan, Ba Dinh, Hanoi.\vspace*{2mm} 
		\\ 
		$^2$Graduate university of science and technology, Vietnam academy of science and technology, 
		18 Hoang Quoc Viet, Cau Giay, Hanoi. 
		\vspace*{2mm} 
		\\
		$^3$Institute for interdisciplinary research in science and education, ICISE, Quy Nhon.
	}
	\email{\\
		$ ^\ast$anhky@iop.vast.ac.vn\\
		$ ^\dagger$phamkyvatly@gmail.com\\
		$ ^\ddagger$nhvan@iop.vast.ac.vn}
	\received{\today}
	\accepted{DD MM YYYY}}
	\maketitle
	\markboth{N. A. Ky, P. V. Ky and N. T. H. Van}{Testing $f(R)$-gravity.}
\begin{abstract}

A procedure of testing the $f(R)$-theory of gravity is discussed. 
The latter is an extension of the general theory of relativity (GR). 
In order this extended theory (in some variant) to be really confirmed  
as a more precise theory it must be tested. To do that we first have 
to solve an equation generalizing Einstein's equation in the GR. 
However, solving this generalized Einstein's equation is often very 
hard, even it is impossible in general to find an exact solution. 
It is why the perturbation method for solving this equation is used. 
In a recent work \cite{Ky:2018fer} a perturbation method was applied 
to the $f(R)$-theory of gravity in a central gravitational field which 
is a good approximation in many circumstances. There, perturbative 
solutions were found for a general form and some special forms of $f(R)$. 
These solutions may allow us to test an $f(R)$-theory of gravity by 
calculating some quantities which can be verified later by the experiment 
(observation). In \cite{Ky:2018fer} an illustration was made on the case 
$f(R)=R+\lambda R^2$. For this case, in the present article, the orbital 
precession of S2 orbiting around Sgr A* is calculated in a higher-order 
of approximation. The $f(R)$-theory of gravity should be also tested for 
other variants of $f(R)$ not considered yet in \cite{Ky:2018fer}. Here, 
several representative variants are considered and in each case the 
orbital precession is calculated for the Sun--Mercury- and the Sgr A*--S2 
gravitational systems so that it can be compared with the value observed 
by a (future) experiment. Following the same method of \cite{Ky:2018fer} 
a light bending angle for an $f(R)$ model in a central gravitational field 
can be also calculated and it could be a useful exercise.\\
\end{abstract}
\section{Introduction}

The General theory of Relativity (GR) is one of the greatest theories of the 20th century.
The heart of the GR is Einstein's equation \cite{SW,LL,P1}
\begin{align} 
R^\mu_{~\nu}-\frac{1}{2}R\delta^\mu_{~\nu}=-\frac{8\pi{G}}{c^4}T^\mu_{~\nu}, 
\label{2}
\end{align}
derived from 
the Lagrangian (Lagrangian density)
\begin{equation}
L_G= R,
\label{elagra}
\end{equation}
where $ T^\mu_{~\nu}$ is the enery-momentum tensor of the matter. 
This theory can explain and predict many gravitational 
phenomena (of the normal matter) in the Universe. Remarkably, recent detections 
of gravitational waves by the LIGO and Vigro collaborations (see, for instance, 
\cite{Abbott:2016blz,TheLIGOScientific:2017qsa}) proved once again the predictive 
power of the GR. If the latter can excellently describe gravitational phenomena of 
the normal matter, it is not, however, a good theory for explanation of a number 
of other phenomena such as dark matter, dark energy, cosmic inflation, etc., as 
well as it cannot accommodate quantum gravity. Various models and theories have 
been suggested to solve these problems. For example, for solving the dark energy 
problem, one of the first and simplest attempts is to add a cosmological constant 
$\Lambda$ to the Lagrangian \eqref{elagra}, becoming $L^\Lambda_G=R-2\Lambda$.
This theory has, however, its own problems (see, for example, \cite{7,DeFelice:2010aj,thomas} 
for more details). \\

One more general but still relatively simple theory\footnote{There are also other models extending the GR, 
however, they are not in the scope of the present paper (see \cite{DeFelice:2010aj,thomas} and references 
therein, for listing some of them).}, expected to solve a wider range of problems in cosmology, is the 
so-called $f(R)$-theory of gravity (or just $f(R)$-theory or $f(R)$-gravity for short) in which Lagrangian 
\eqref{elagra} is replaced by 
\begin{align}
{\cal L}_G= f(R),
\label{flagra}
\end{align}
which is a scalar function of the scalar curvature $R$. Thus, Einstein's equation must be replaced by the 
equation \cite{DeFelice:2010aj, thomas, Capozziello:2011et}
\begin{align}
f'(R)R^\mu_{~\nu}-\delta^\mu_{~\nu}\square f'(R)+\nabla^{\mu}\nabla_{\nu} f'(R)
-\frac{1}{2}f(R)\delta^\mu_{~\nu}=-kT^\mu_{~\nu}, 
\label{feq}
\end{align}
where $k= \frac{8\pi G}{c^4}$, $ \square = \nabla_{\mu}\nabla^{\mu}$ with $\nabla_{\mu}$ being 
a covariant derivative and $ f'(R)=\frac{df(R)}{dR} $. Presently, the $f(R)$-theory 
is one of the hotest topics in cosmology with 
different versions of $f(R)$ considered (for review, see, for example, \cite{DeFelice:2010aj} -- 
\cite{Capozziello:2009vr}) 
such as those with $ f(R)=R+\lambda R^2 $ or $ f(R)=R-\frac{\lambda}{R^n} $, etc.
However, to solve Eq. \eqref{feq}, especially, for an exact solution, is usually very difficult, 
even impossible. To get rid of this situation, some approximation conditions are sometimes required 
so that approximate solutions can be found. Among such conditions the spherical symmetry which 
is a quite good approximation in many cases is often chosen. Following this strategy in a recent work 
\cite{Ky:2018fer} we solved Eq. \eqref{feq} for a general $f(R)$-theory in a central (gravitational) 
field which in general is not static, and obtained approximate solutions in vacuum and in the presence 
of matter. Then, as a test and illustration, applications of these solutions for $f(R)=R+\lambda R^2$ 
are presented. In the present paper we continue testing other versions of the $f(R)$-theory. Before 
doing that in Sect. 3, we will briefly recall in the next section some results of \cite{Ky:2018fer} 
to make this paper more self-contained. Some conclusions and comments are given in the last section, 
Sect. 4. Here, for convenience, we keep the conventions used in \cite{Ky:2018fer}.\\
\section{Perturbative solutions of the $f(R)$-theory in a central field}

Let us summarize some results obtained in \cite{Ky:2018fer}. As the GR is a very precise theory, 
it is reasonable to assume that a realistic $f(R)$ theory differs just slightly from the GR, 
that is, $f(R)$ can be written in the form 
\begin{align}
f(R)=R+\lambda h(R) \label{flambda},
\end{align}
where $\lambda$ is a parameter and $h(R)$ is a scalar function of $R$ such that $\lambda h(R)$ 
and its derivatives are very small quantities in comparison with $R$. 
With $f(R)$ given in \eqref{flambda} the modified Einstein's equation \eqref{feq} becomes  
\begin{align}
R^{\mu}_{~\nu}-\frac{1}{2}\delta^{\mu}_{~\nu}R+\lambda h'(R)R^{\mu}_{~\nu}
-\frac{\lambda}{2}\delta^{\mu}_{~\nu}
h(R)-\lambda \delta^{\mu}_{~\nu}\square h'(R) 
+\lambda \nabla^{\mu}\nabla_{\nu}h'(R)=-kT^{\mu}_{~\nu}. \label{8}
\end{align}
Solving this equation for a central field of a gravitational source of mass 
$M$ we obtain a Schwarzschild-type solution ($ x^0=ct $)
\begin{align}
ds^2=\left[ 1-\frac{2GM_f(t)}{c^2r}\right] {dx^0}^2-\frac{dr^2}{1-\frac{2GM_f(t)}{c^2r}}
-r^2(d\theta^2+\sin^2\theta{d\varphi^2}) \label{a81}
\end{align}
with
\begin{align}
M_f(t)=M-\lambda M_1(t)-\lambda M_2(t), \label{a80}
\end{align}
treated as an effective mass, which in general is a function of time, even for a constant $M$, 
where
\begin{align}
&\lambda M_1(t)=\frac{2\pi\lambda [R_0(t)]^3}{3kc^2}\left[h(kT^0_{~0})+kT^0_{~0}h'(kT^0_{~0})\right],  \label{a74}\\
&\lambda M_2(t)=\frac{4\pi\lambda}{kc^2}h''(kT^0_{~0})\left[ \frac{\partial}{\partial t}
\frac{M}{[R_0(t)]^3}\right]^2 ~\alpha (t),  \label{a75}
\end{align}
\begin{align} 
T^0_{~0}= {M c^2\over {\frac{4}{3}}\pi {[R_0(t)]}^3},
 \label{24}
\end{align}
\begin{align}
\alpha (t)= ~
&
\frac{3k^2c^2R_0(t)}{256\pi^2[\xi (t)]^4}\left\lbrace \frac{3}{\xi(t)R_0(t)}
\arcsin[\xi (t) R_0(t)]
-\left( 3+2[\xi(t)R_0(t)]^2\right) \sqrt{1-[\xi(t)R_0(t)]^2}\right\rbrace
\nonumber \\
&
\times \left( 1-[\xi (t)R_0(t)]^2\right)^{-3/2}, \label{37}
\end{align}
and
\begin{align}
\xi^2 (t)=\frac{kMc^2}{4\pi [R_0(t)]^3}. \label{37a}
\end{align}
Above, the radius of the considered body-gravitational source $R_0$ is also 
a function of time, $R_0 = R_0(t)$, in general. 
If the body-gravitational source shrinks or expands (it means 
that its radius varies with time), the  metric would depend on time. 
This affect does not happen in the GR and may lead to new phenomena 
which is a subject of our current research.\\

\begin{figure}[h]
\begin{center}
\includegraphics[scale=0.2]{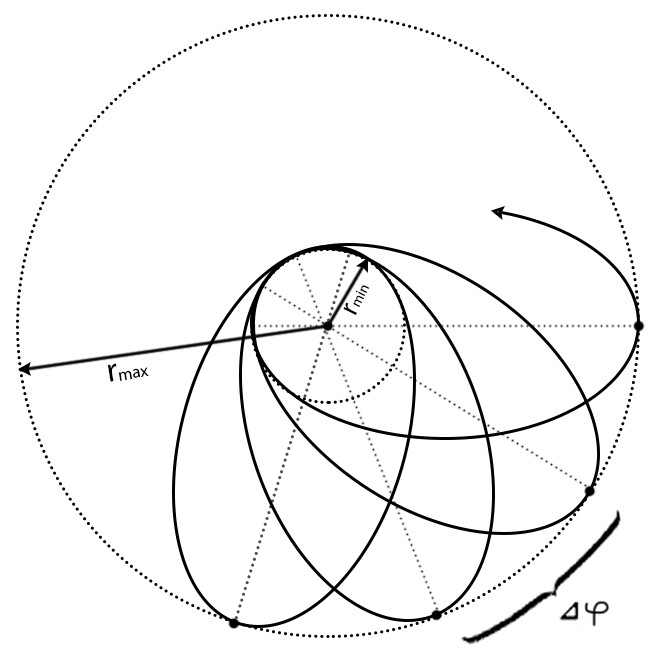}
\caption{\label{fig:1}\textit{In a static central field, both  $r_{p/a}$ and  $ \Delta \varphi $ 
remain constant as in the GR but differ from the corresponding Einstein's 
values by constant corrections}\cite{Ky:2018fer}.}
\end{center}
\end{figure}

Applying the solution \eqref{a81} to the problem of a planet orbiting 
around an isotropic star of mass $M$ we find the equation of motion 
\begin{align}
\frac{l^2(t)}{m\beta(t)r}=1+\sqrt{1+\frac{2E(t)l^2(t)}{m\beta^2(t)}}\mbox{cos}
\left( \sqrt{1-\frac{6m^2G^2M_f^2(t)}{c^2\mu^2}}\varphi\right), \label{a109}
\end{align}
with $r$ and $\varphi$ being polar coordinates of the planet in a frame with origin at 
the star's center, while 
\begin{align}
\beta(t)= ~& mGM_f(t)\left[ 1+\frac{4E(t)}{mc^2}\right],  \label{a104}\\
l^2(t)= ~& \mu^2\left[ 1-\frac{6m^2G^2M_f^2(t)}{c^2\mu^2}\right],  \label{a105}
\end{align}
where $E(t)$ is the energy of the planet (subtracted by the rest energy $mc^2$) in the 
gravitational field, and $\mu$ is  the angular momentum (which is conserved). The orbit 
described by \eqref{a109} is nearly-elliptic with parameters, such as major and minor axes, 
changing with time if the central field is not static (even when the total mass $M$ is constant, 
as, for example, in the case of a start expanding or collapsing but keeping its isotropic form).
Following \cite{Ky:2018fer} we can calculate the minimal value $r_p$ and the maximal value 
$r_a$ of $r$ 
\begin{align}
r_{p/a}=&\frac{l^2(t_{e})}{m\beta(t_{e})\pm \sqrt{m^2\beta^2(t_{e})+2mE(t_{n})l^2(t_{e})}}, \label{a112}
\end{align}
where the signs plus and minus are for $r_p$ and $r_a$, respectively,  and 
$ t_e $ is the time at the extremum $r_e$ being $r_p$ or $r_a$.
The orbital precession can be also calculated 
\begin{align}
\Delta \varphi_e (k)=\frac{6\pi m^2G^2M_f^2(t_k)}{c^2\mu^2}.\label{Detaphi}
\end{align} 
The latter differs from Einstein's precession by a correction 
which at the first order of perturbation reads
\begin{align}
\delta \varphi_e (k) \cong 
\frac{-12\pi m^2G^2\lambda M[M_1(t_k)+M_2(t_k)]}{c^2\mu^2},
\label{deltaphi2}
\end{align}
where $ t_k $ ($ k=1, 2, 3, ...$) is the time when the planet passes the extremum points $r_k$, 
which are either $r_p$ or $r_a$ (but not both). 
If the central field is static, $\Delta \varphi_e$ (therefore, $\delta \varphi_e$) remains always 
constant (see Figure \ref{fig:1} for illustration), 
but when the central field is not static, $\Delta \varphi_e$ (therefore, $\delta \varphi_e$) 
may change with time.
\begin{figure}[h]
\begin{center}
\includegraphics[scale=0.2]{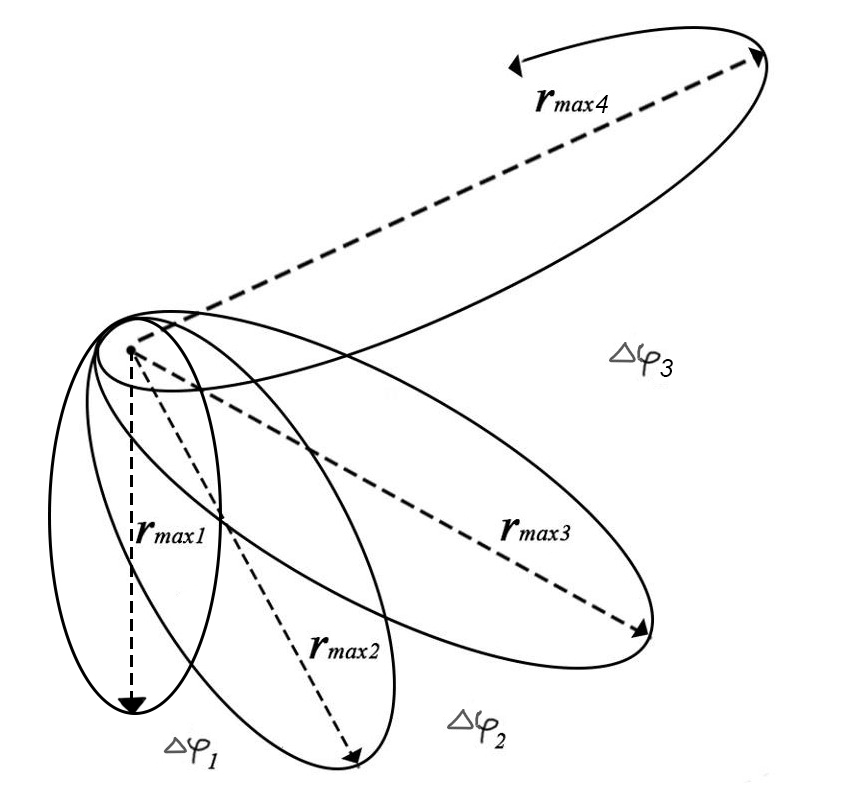}
\caption{\label{fig:2}\textit{In a non-static case, both $r_{p/a}$ and $\Delta \varphi$ 
in general vary with time, unlike in the GR they remain always constant in a central 
field (if the source has a constant mass) \cite{Ky:2018fer}}.}
\end{center}
\end{figure}
There is not only a correction \eqref{Detaphi} to the orbital precession, 
but, as seen from \eqref{a112}, the orbital axes also change with time 
(for illustration, see Figure \ref{fig:2}). They are new effects compared 
with the GR and require to be tested. In \cite{Ky:2018fer} testing the 
$f(R)$-theory was illustrated with $f(R)=R+\lambda R^2$, here, in this 
paper, we will do that with other variants.  

\section{Testing the $f(R)$-theory in some variants}

For simplicity, let us consider an $f(R)$-theory in a static central field. 
As seen before, our perturbative solution looks like a Schwarzschild solution 
in the GR with the original mass $M$ replaced by an effective mass 
$M_f=M-\lambda M_1-\lambda M_2$ which now is just $M_f=M-\lambda M_1$ because 
$M_2=0$ for a static field. Then, from \eqref{Detaphi} we have 
\begin{equation}
\Delta \varphi_{f(R)}
=\frac{6\pi G^2 m^2 M_{f}^2}{c^2 \mu^2}
= \frac{6\pi G^2 m^2 (M-\lambda M_1)^2}{c^2 \mu^2}. \label{H1}
\end{equation}
This latter can be written in the form 
\begin{equation}
\Delta \varphi_{f(R)}=\frac{6\pi G(M-\lambda M_1)}{c^2a(1-e^2)}, \label{H1ab}
\end{equation}   
where 
\begin{equation}
\frac{\mu^2}{GM_fm^2}=a(1-e^2) \label{H1aa}
\end{equation}
is used, with $a$ being the length of a semi-major axis and $e$ being the eccentricity 
of an orbital ellipse \cite{LL}, hence
\begin{equation}
\lambda M_1=M-\frac{c^2a(1-e^2)}{6\pi G}\Delta \varphi_{f(R)}. \label{H12}
\end{equation} 
It is worth noting that this formula, valid for any (well-defined) $f(R)$, 
not only for $f(R)=R+\lambda R^2$, is different from (149) in \cite{Ky:2018fer} because 
it is calculated in a higher order of precision by using in \eqref{H1aa} the effective 
mass $ M_f=M-\lambda M_1 $ instead of the "bare" mass $M$ used in \cite{Ky:2018fer}. 
The value of  $\lambda M_1$ is the same for an arbitrary $f(R)$ but $M_1$, thus $\lambda$, 
is different for different $f(R)$. Next, using the following data \cite{Majumder:2011eh}: 
\begin{align}
&c=299792458 m/s; \nonumber\\
&G=6.67259\times 10^{-11} kg^{-1}m^3s^{-2};\nonumber\\
&\frac{2GM}{c^2}=2.95325008\times 10^3 m; \nonumber\\
&k=\frac{8\pi G}{c^4}=2.0761154\times 10^{-43} kg^{-1}m^{-1}s^2;\label{H2a}\\
&M=1.988919\times 10^{30}kg;\nonumber\\
&a=5.7909175\times10^{10}m; \nonumber\\
&e=0.20563069;\nonumber\\
&\Delta\varphi_{obs}=2\pi(7.98734\pm 0,00037)\times 10^{-8}~radian/revolution, \nonumber
\end{align} 
we obtain the deviation between the observed value and the GR value of the Mercury's 
orbital precession
\begin{align}
\Delta \varphi_{obs}-\Delta \varphi_{GR}
=  -0.1906\pi \times 10^{-11} radian/revolution.
\label{phiSun1}
\end{align}
This deviation may come from the imperfection, though small, of the GR, and suppose it 
can be explained by the $f(R)$- theory, that is, 
\begin{equation}
\Delta \varphi_{f(R)}= \Delta \varphi_{obs}  \label{phiSun2}
\end{equation} 
(upto some smaller errors),  
or
\begin{align}
\delta\varphi= & ~\Delta \varphi_{f(R)}-\Delta \varphi_{GR}
=  -0.1906\pi \times 10^{-11} radian/revolution. 
\label{phiSun}
\end{align}
This requirement can be satisfied if the correction $\lambda M_1 $ to the mass $M$ equals   
\begin{equation}
\lambda M_1=23.285244\times10^{24}kg, \label{H3}
\end{equation} 
according to \eqref{H12} and \eqref{phiSun2}. The effective reduction of the Sun's mass  
$ M=1.988919\times 10^{30}kg $ 
\begin{equation}
\frac{\lambda M_1}{M}=11.707487\times 10^{-6}=0.0011707487 ~\%. \label{H4}
\end{equation} 
is quite small compared with the Sun's mass but it may be measurable if a measurement 
technique precise enough is invented. It is worth noting that the value of $\lambda M_1$ 
is model-independent, i.e., for any well-defined $f(R)$. To estimate $\lambda$ we need, 
however, a concrete $f(R)$. \\ 

Using the perturbation condition $\lambda h(R)\ll R$ for \eqref{flambda} and $R=kT$, 
where $T=T^\mu_\mu$, 
we get   
\begin{equation}
\lambda h(kT)\ll kT, \label{H5}
\end{equation}
or if $T\approx T^0_{~0}$, we have 
\begin{equation}
\lambda h(kT^0_{~0})\ll kT^0_{~0}. \label{H6}
\end{equation}
With $T^0_{~0}=\frac{Mc^2}{\frac{4}{3}\pi [R_0]^3}$  the latter inequation becomes 
\begin{equation}
\lambda h(kT^0_{~0})\ll \frac{6GM}{c^2[R_0]^3}. \label{H7}
\end{equation}
From this formula with the Sun's radius \cite{Haberreiter:2007ku}
\begin{equation}
R_0\approx 6.957 \times 10^8 m \label{ridius Sun}
\end{equation}
we see that $\lambda h$ is very small, 
\begin{equation}
\lambda h(kT^0_{~0})\ll 26.3120915 \times 10^{-24}, \label{H7b}
\end{equation}
as expected. All results listed above are for an arbitrary $f(R)$. 
Now let us consider several concrete variants of the $f(R)$-theory.\\

\subsection{Model $ f(R)=R+\lambda R^2 $ \label{subsec1}}
 
 This  called Starobinsky model \cite{Starobinsky:1980te} was already 
considered in \cite{Ky:2018fer} (more discussions on the meaning of 
this model can be found in \cite{DeFelice:2010aj}) as model II with $b=2$, but here 
we reconsider it by doing some calculations at a higher order of precision, 
namely, as said above, formula \eqref{H1aa} with $M_f$ replacing $M$ is used instead of 
$\frac{\mu^2}{GMm^2}=a(1-e^2)$ used in \cite{Ky:2018fer}. In this model $h(R)=R^2$, that 
is $h(kT^0_{~0})=[ kT^0_{~0}]^2$, and, therefore, due to  \eqref{24} we have   
\begin{equation}
h(kT^0_{~0})=\left[ \frac{kMc^2}{\frac{4}{3}\pi [R_0]^3}\right] ^2, \label{T1}
\end{equation}
and  
\begin{equation}
kT^0_{~0}h'(kT^0_{~0})=2\left[ \frac{kMc^2}{\frac{4}{3}\pi [R_0]^3}\right] ^2. \label{T2}
\end{equation}
Thus, the perturbation condition \eqref{H7} for this model imposes an upper bound  on $\lambda$:  
\begin{equation}
\lambda \ll \frac{c^2[R_0]^3}{6GM}. \label{T3}
\end{equation}
Using the data in \eqref{H2a} and \eqref{ridius Sun} we can calculate this bound, 
\begin{equation}
\lambda \ll 0.380053\times 10^{23}. \label{T4}
\end{equation}
Now inserting \eqref{T1} and \eqref{T2} in \eqref{a74} we write $M_1$ in the form    
\begin{equation}
M_1=\frac{9kc^2M^2}{8\pi[R_0]^3},  \label{T5}
\end{equation}
which, with using \eqref{H2a} and \eqref{ridius Sun} again, gives  
\begin{equation}
M_1=78.4989\times 10^6 kg. \label{T6}
\end{equation}
From here and \eqref{H3} we obtain a numerical value of $\lambda$,   
\begin{equation}
\lambda=0.296631\times 10^{18}. \label{T7}
\end{equation}
The latter is compatible with the perturbation condition \eqref{T4}. 
The model $f(R)=R+\lambda R^2$ with $\lambda$ given in \eqref{T7} 
makes a small correction to the GR and fits the observed Mercury's 
orbital precession. Now, assuming that the obtained value of $\lambda$ 
is universal (at least within some range of gravitational field strength) 
we can predict an orbital precession for another gravitational system. 
Following this procedure described in more details in \cite{Ky:2018fer} 
and using the data \cite{Gillessen:2008qv}
\begin{align}
&M=4.31\times 10^6 M_\odot=8.57\times 10^{36} kg \nonumber\\
&R_0=22\times 10^{9}m  \label{T8}\\
&a=0.123arcsec =14.7\times 10^{13}m \nonumber\\
&e=0.88, \nonumber
\end{align}
we can calculate an improved S2 orbital precession 
$\Delta\varphi^{S2}_{f(R)}$ around SgrA* as  
\begin{equation}
\Delta\varphi^{S2}_{f(R)}=1.149305\pi\times 10^{-3}radian/revolution. \label{phiSgrA2}
\end{equation}
This value of $\Delta\varphi^{S2}_{f(R)}$ slightly improves the one  
calculated in \cite{Ky:2018fer} and its deviation from the GR's value is 
a bit bigger, thus, more measurable. Now we are moving to other models not 
considered in details yet in the previous work \cite{Ky:2018fer}, but 
we should note first that any function $f(R)$ which can develop a Taylor expansion 
around $R=0$, coincides at the leading order with $f(R)=R+\lambda R^2$.\\

\subsection{Model $f(R)=R+\lambda R^2\sum\limits_{n=0}^{+\infty}{a_nR^n} $  \label{subsec2} }

This model is inspired by the Taylor expansion of $f(R)$ considered also in 
\cite{Psaltis:2007cw, Berry:2011pb} stating that an $f(R)$ theory can be distinguished with 
the GR only beyond a Kerr solution.
Here $a_n$ is a coefficient regulating a right dimension of each term $a_nR^n$, where, $a_0$ 
can be normalized to be 1, $a_0=1$.
As according to \eqref{H7b} 
\begin{equation}
R=kT^0_{~0}= 26.3120915 \times 10^{-24} \label{T9}
\end{equation}
for the Sun and 
\begin{equation}
R=kT^0_{~0}=35.8523036\times 10^{-22} \label{T10}
\end{equation}
for the SgrA*, i.e., $R\ll 1$ in both cases, this model \eqref{subsec2} is convergent 
if $a_n$ are not very big.
As $R\ll 1$ the approximation $f(R) \approx R+\lambda R^2$ can be used, and the investigation 
of this model is similar to that of the previous one \eqref{subsec1}. We classify those models 
with the same lower-order approximation into one class. The next model also belongs to this class. \\

\subsection{Model $ f(R)=Re^{\lambda R} $  \label{subsec3}}
Similar to the model considered above, the present model is also inspired by the Taylor 
expansion of an $f(R)$ for a special choice of coefficients $a_n$
(see \cite{Linder:2009jz} for another resembling model).
One can see from the Taylor series
\begin{equation}
f(R)=Re^{\lambda R}=R\sum\limits_{n=0}^{+\infty}\frac{(\lambda R)^n}{n!}=R+\lambda R^2
+\frac{\lambda^2R^3}{2}+\frac{\lambda^3R^4}{3!}+\dotsb \label{T12}
\end{equation}
that this model belongs to the same class with the models \eqref{subsec1} and \eqref{subsec2}. 
These models describe the same physics, including the same $\lambda$, at the order $R^2$ 
of approximation of $f(R)$. There are other models belonging to this class but we cannot 
list all here. Of course, if we go to a higher order of approximation we have to do more 
cumbersome, sometimes,  impossible, calculations but the general procedure remains the 
same. As the correction to the GR is already very small even at the leading order of 
approximation there is no need at the present to make calculations at the next orders. 
Therefore, all the $f(R)$ models with $f(R)$ developing a Taylor expansion 
around $R=0$, describe at the leading order of approximation the same physics as Starobinsky 
model, that is, they belong to one and the same class referred hereafter to as Starobinsky class. 
Below we will consider some alternative models, not equivalent to Starobinsky model.\\

\subsection{Model $f(R)=\alpha R^{1+\varepsilon}$  \label{subsec4}}

 Here $\varepsilon$ is an infinitesimally small number and $\alpha$, which may 
depend on $\varepsilon$, is a coefficient regulating a right dimension of $f(R)$. 
To get Einstein's GR at $\varepsilon =0$ requires 
$\lim\limits_{\varepsilon \rightarrow 0}\alpha (\varepsilon)=1$. 
Perturbative solutions of the current model were already considered 
in \cite{Ky:2018fer} but here we suggest a testing procedure. 
This model for $\varepsilon<0$ (as seen below) does not belong to 
Starobinsly class as the function $f(R)=\alpha R^{1+\varepsilon}$ cannot develop 
a Taylor expansion around $R=0$ (but it could belong to a class of a type with a 
cosmological constant).
Now we have      
\begin{equation}
\lambda h(R) = \alpha R^{1+\varepsilon}-R, \label{T13}
\end{equation}
and, thus,  
\begin{equation}
\lambda Rh'(R) = (1+\varepsilon )\alpha R^{1+\varepsilon}-R. \label{T14}
\end{equation}
Inserting \eqref{T13} and \eqref{T14} in \eqref{a74} we get 
\begin{equation}
\lambda M_1=-M+\alpha \frac{M}{2}(\varepsilon +2)\left(\frac{3kc^2M}
{4\pi [R_o]^3}\right)^\varepsilon \label{T15}
\end{equation}
and then combining \eqref{T15} with \eqref{H3} we obtain the equation 
\begin{equation}
-M + \alpha \frac{M}{2}(\varepsilon +2)\left(\frac{3kc^2M}{4\pi [R_o]^3}\right)^\varepsilon
=23.285244\times10^{24}. \label{T16}
\end{equation}
Using the data from \eqref{H2a} we solve this equation for $\varepsilon$ to get  
\begin{equation}
\varepsilon=-2.27364\times 10^{-7}, \label{T17}
\end{equation}
and from here with \eqref{T9} and \eqref{T13} we find $\lambda h(R)=31.1040024\times 10^{-29}$. 
This value of $\lambda h(R)$ also satisfies the perturbation condition $\lambda h(R)\ll R$. 
With the value of $\epsilon$ given in \eqref{T17} we can calculate an orbital precession 
of Mercury fitting the observed one with the correction \eqref{phiSun} to the Einstein 
value as we did earlier for other models. \\

In this way we can calculate the orbital precession of S2 orbiting around our Galaxy's center 
Sgr A*. Taking \eqref{T17}, \eqref{T15}, \eqref{T8} and \eqref{H1ab} into account we get 
\begin{equation}
\Delta\varphi_{f(R)}^{S2}=1.15113\pi\times 10^{-3} radian/revolution, \label{T18}
\end{equation}
and the correction to the Einstein value is 
\begin{equation}
\delta\varphi^{S2}=\Delta\varphi_{f(R)}^{S2}-\Delta\varphi_{GR}^{S2}
=-\pi\times 10^{-8} radian/revolution, \label{T19}
\end{equation}
where 
\begin{equation}
\Delta\varphi^{S2}_{GR}=
1.15114\pi\times 10^{-3}radian/revolution \label{phiSgrA1}
\end{equation}
is the orbital precession calculated by the GR.\\

\subsection{Model $ f(R)=R+\lambda\sqrt{R} $  \label{subsec5}}
Assuming $ \lambda \sqrt{R}\ll R $, the present model 
is inspired by the model $ f(R)=R+\lambda R^n $ with $n>0$ (see, for 
example, \cite{DeFelice:2010aj}). This model does not 
belong to Starobinsky class (but a class with a cosmological-type  
constant) either, because $f(R)=R+\lambda\sqrt{R}$ cannot develop a 
Taylor expansion around $R=0$.
In this case  
\begin{equation}
h(R)=\sqrt{R}, \label{T20}
\end{equation}
\begin{equation}
Rh'(R)=\frac{\sqrt{R}}{2}. \label{T21}
\end{equation}
Now with \eqref{T20} the perturbation condition \eqref{H7} becomes  
\begin{equation}
\lambda\ll \sqrt{kT^0_{~0}}=5.129531314\times 10^{-12}. \label{T22}
\end{equation}
Inserting \eqref{T20} and \eqref{T21} in \eqref{a74} we obtain the 
formula 
\begin{equation}
M_1=\frac{\sqrt{3\pi M}[R_0]^{\frac{3}{2}}}{2c\sqrt{k}}, \label{T23}
\end{equation}
which with \eqref{ridius Sun} and \eqref{H2a} taken into account gives 
\begin{equation}
M_1=2.908041992\times 10^{41}. \label{T24}
\end{equation}
Combining \eqref{T24} with \eqref{H3} we get  
\begin{equation}
\lambda=8.007189739\times 10^{-17}. \label{T25}
\end{equation}
We see that this value of $\lambda$ satisfies the perturbation condition \eqref{H3} 
securing the orbital precession of Mercury calculated by the present model fits 
the observed one, and, therefore, it makes a correction to the Einstein value as 
given by \eqref{phiSun}. Following the same procedure we can calculate the orbital 
precession of S2 around Sgr A* 
\begin{equation}
\Delta\varphi_{f(R)}^{S2}=1.15114\pi \times 10^{-3}radian/revolution, \label{T26}
\end{equation}
which is (almost) the same as the value \eqref{phiSgrA1} obtained by the GR. \\

\section{Conclusions}

The general theory of relativity is very successful theory which is the foundation of 
the modern cosmology, but it cannot solve a number of problems like those of dark matter, 
dark energy, inflation, quantum gravity, etc., that require a modification or an extension 
of this theory. The so-called $f(R)$-theory of gravity is one of the most popular and 
simplest modified theories of gravity generalizing the GR in order to resolve difficulties 
of the latter. As any other new theory the $f(R)$-theory must be verified by the experiment 
(observation). One of the ways to do that is to compare some theoretically derived quantities 
with the corresponding measured (observed) ones. Therefore, we have to prepare theoretical 
samples to be checked later experimentally. In the present article, using the method of 
\cite{Ky:2018fer} we have calculated  for several representative variants of the $f(R)$-theory 
orbital precessions, which should be compared with the available measured values. Following 
\cite{Ky:2018fer} it is not difficult to calculate a light bending angle for an $f(R)$ model 
in a central gravitational field, but here we have calculated orbital precessions as examples and 
leave the calculations on the light bending as an exercise for those interested. We hope a 
precision measurement of these quantities can be organized in a not very far future. 
For conclusion, we have considered for testing several variants of the 
$f(R)$-gravity, but so far, before having experimental/observation data, we cannot 
compare them in order to say which is a better, i.e., more realistic, model. However, 
we can state that the orbital precessions calculated by the models of the class III.1 -- III.3 
deviate more from the GR than those for III.4 and III.5, thus, the former are easier to be 
experimentally tested. Some of the above considered models have found physical 
interpretations (to be verified experimentally) but other ones suggested just as alternative 
possibilities within a mathematical completion may find later physical interpretations.
\section*{Acknowledgement} 

    This research is funded by Vietnam's National Foundation for Science and 
Technology Development (NAFOSTED) under contract No 103.01-2017.76.

\end{document}